\documentclass{article}
\usepackage[preprint]{spconf}
\usepackage{amsmath,graphicx,dirtytalk,bm,amsfonts,acronym,booktabs,multirow,balance}
\usepackage[caption=false,font=normalsize]{subfig}
\usepackage[T1]{fontenc}
\usepackage[hidelinks,implicit=false]{hyperref}
\usepackage[capitalise]{cleveref}
\newcommand{\ev}[1]{\mathbb{E}[#1]}

\newcommand{\vect}[1]{\bm{#1}}

\newcommand*{\tpose}{^\mathsf{T}}

\newcommand*{\reals}{\mathbb{R}}

\newcommand*{\naturals}{\mathbb{N}}
\newcommand*{\complexes}{\mathbb{C}}
\newcommand*{\iu}{j}

\acrodef{STFT}{short-time Fourier transform}
\acrodef{MSE}{mean-squared error}
\acrodef{MAE}{mean absolute error}
\acrodef{SUB}{subtractive}
\acrodef{CSE}{channel separation estimate}
\acrodef{SOTA}{state-of-the-art}
\acrodef{AR}{augmented reality}
\acrodef{NS}{noise suppression}
\acrodef{SS}{reverberant speech separation}
\acrodef{DR}{de-reverberation}

\toappear{To appear in {\it Proc.\ ICASSP2023, June 04-10, 2023, Rhodes Island, Greece.}}

\title{Towards Real-Time Single-Channel Speech Separation in Noisy and Reverberant Environments}

\twoauthors
 {Julian Neri \sthanks{Performed work during an internship at Microsoft Research.}}
	{McGill University, CIRMMT, Montreal, CA\\
	\url{julian.neri@mcgill.ca} }
 {Sebastian Braun}
	{Microsoft Research, Redmond, USA \\
	\url{sebastian.braun@microsoft.com}}

\begin{document}
\ninept
\maketitle
\begin{abstract}
Real-time single-channel speech separation aims to unmix an audio stream captured from a single microphone that contains multiple people talking at once, environmental noise, and reverberation into multiple de-reverberated and noise-free speech tracks, each track containing only one talker. While large state-of-the-art DNNs can achieve excellent separation from anechoic mixtures of speech, the main challenge is to create compact and causal models that can separate reverberant mixtures at inference time. In this paper, we explore low-complexity, resource-efficient, causal DNN architectures for real-time separation of two or more simultaneous speakers. A cascade of three neural network modules are trained to sequentially perform noise-suppression, separation, and de-reverberation.  For comparison, a larger end-to-end model is trained to output two anechoic speech signals directly from noisy reverberant speech mixtures. We propose an efficient single-decoder architecture with “subtractive” separation for real-time recursive speech separation for two or more speakers. Evaluation on real monophonic recordings of speech mixtures, according to speech separation measures like SI-SDR, perceptual measures like DNS-MOS, and a novel proposed channel separation metric, show that these compact causal models can separate speech mixtures with low latency, and perform on par with large offline state-of-the-art models like SepFormer.
\end{abstract}
\section{Introduction}
\label{sec:intro}

Separating multiple overlapping speech signals from a single microphone signal is a difficult problem that has many applications like automatic dictation and meeting transcription, \ac{AR}, and hearing aids. 
Due to single-channnel speech separation's relevant practical applications and theoretical links to human perception of auditory scenes, namely the \say{cocktail party effect} \cite{Darwin:2007:Speech:Segregation:Cocktail}, it has been a primary topic of signal processing research for several decades \cite{Wang:2018:Speech:Separation:DNN:Overview}. While research in speech enhancement and separation has made tremendous progress in the past few years, \ac{SOTA} models still are challenged in realistic acoustic environments due to generalization issues, and do not fulfill low-latency processing requirements in terms of look-ahead and computational complexity.

Large deep neural networks (DNNs) like ConvTasNet \cite{Luo:2019:ConvTasNet}, dual-path RNN \cite{Luo:2020:Dual:Path:Rnn:Speech:Separation}, and SepFormer \cite{Subakan:2022:Transformer:Speech:Separation} enable state-of-the-art speech separation quality, far outperforming traditional statistical approaches \cite{Vincent:2018:Source:Separation}, due to their higher modeling capacity and straightforward end-to-end optimization.
However, so far all \ac{SOTA} DNNs are not practicable in real-time settings as they are not resource efficient, non-causal, and delay the output significantly due to their complexity and dependency on a large look-ahead (future input values).
%
Another major challenge is the generalization of data-driven separation algorithms to real world conditions. Most papers report results only on synthetic and even ancechoic conditions, and therefore observe large degradations on realistic data, i.e.\,in noisy and reverberant conditions \cite{Maciejewski:2020:Whamr:dataset:reverb,Landwehr:2021:Reverberant:Separation:Sepformer}. 
Experiments in \cite{Maciejewski:2020:Whamr:dataset:reverb} showed that a cascade of task-specific networks performed better than a single network in terms of signal-to-distortion ratio given noisy reverberant speech mixtures.


In this paper, we develop resource-efficient and causal DNN architectures for real-time separation of noisy and reverberant single-channel speech mixtures that are an order of magnitude cheaper in complexity than current \ac{SOTA} DNNs used for speech separation, while having a small look-ahead (system latency).
We show that our cascaded three-stage system consisting of noise suppression, speech separation, and de-reverberation modules outperforms our single-stage baselines, both in terms of overall estimation quality and computational complexity, and is competitive in separated speech quality with non-real-time large DNNs especially when the input mixtures have significant room reverberation and noise.
We propose a novel single-output separation module with \ac{SUB} separation and compare it to multi-output separation modules, and find that \ac{SUB} is more resource efficient without sacrificing quality, and is easily extended to separate arbitrary speakers in a recursive way like \cite{2018:Kinoshita:Subtractive:Recurrent:Hearing:Nets,Takahashi:2019:Recursive:Speech:Separation:unknown}.
Investigating the estimation quality of quieter secondary talkers, we combine a perceptually inspired compressed loss with soft thresholding \cite{Wisdom:2020:Mixture:Invariant:Unsupervised}.
Compared to the current \ac{SOTA} scale-invariant loss \cite{Roux:2019:Si:SDR}, we show that our proposed compressed complex loss can give better performance with no additional cost.
Audio examples and supplemental results are available at \url{https://www.music.mcgill.ca/~julian/real-time-speech-separation}.

\section{Problem formulation}
A mixture of $R$ reverberant speech signals and noise is given by
\begin{align}
    y(t) = \eta(t) + \sum_{r=1}^R s_r(t) \ast h_r(t) \, ,
\end{align}
where $\eta(t)$, $s_r(t)$, and $h_r(t)$  are the noise signal, the $r^\text{th}$ anechoic source signal, and the room impulse response (RIR) from the $r^\text{th}$ source to the microphone, respectively, and $t$ is the time index.

The goal is to recover $R$ source signals, $s_r(t)$, from a single channel recording of the mixture, $y(t)$. This involves solving a conflation of three problems: noise suppression (NS), speech separation (SS), and speech de-reverberation (DR).

The \ac{STFT} of a time-domain signal is denoted with capital letters, e.g.\,$Y(k,n) \in \complexes$ is the STFT of $y(t)$ evaluated at frequency index $k \in \naturals$ and time index $n \in \naturals$.

\section{Modular system architecture}
\label{sec:method}

In this work, we propose causal separation systems that only require the features of the current and several past frames to infer the current frame.
The algorithmic delay (latency) of the systems depends only on the \ac{STFT} window and hop size. This allows use of the systems for real-time applications like live communication or \ac{AR}.
We propose a modular system design, with dedicated blocks for each task,  \ac{NS}, \ac{SS}, and \ac{DR}. We use the same network module as described in Sec.~\ref{ssec:general:model} for each module, and follow the order also found to perform best in \cite{Maciejewski:2020:Whamr:dataset:reverb}, inverting the physical creation process.

\subsection{General model architecture}
\label{ssec:general:model}

\begin{figure}[t!]
    \centering
    \includegraphics[scale=.8]{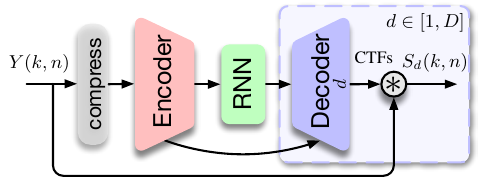}
    \caption{General model architecture. The end-to-end model and separation module involve multiple decoders ($D>1$), whereas modules for noise suppression and de-reverberation use one decoder ($D=1$).}    
    \label{fig:cruse:2out}
\end{figure}

The networks are based on a generalized version of the Convolutional Recurrent U-net for Speech Enhancement (CRUSE) architecture \cite{Braun:2021:Real:Time:Noise:Suppression:CRUSE}, as shown in \cref{fig:cruse:2out}.
The CRUSE model uses 4 convolutional encoder layers, mirrored transposed convolutional decoder layers, causal two-dimensional kernels of size $(2,3)$ for time and frequency dimensions, and downsampling along the frequency dimension using strides $(1,2)$.
The encoder takes as input the complex compressed versions of \ac{STFT} $Y$.
A recurrent network sits between the encoder and decoder to integrate temporal information. 
From each encoder layer, a skip connection with $1\times 1$ convolution is added to the input of each corresponding decoder layer.

We generalize CRUSE to have $D \geq 1$ decoders that independently output convolutive transfer functions (CTFs) \cite{Talmon:2009:CTF:Transfer:Function} also known as deep filters \cite{Mack:2020:Deep:Filtering}, as we found these to perform better than the common approach with real or complex time-frequency masks.
In parallel, each decoder's CTFs filter the \ac{STFT}-domain input $Y$ to produce $S_d$, $\forall d \in [1,D]$.
We consider either GRU or LSTM bottleneck layers, with one or multiple sequential layers.

\subsection{Cascaded three-stage model}

\begin{figure}[t!]
    \centering
    \includegraphics[scale=.8]{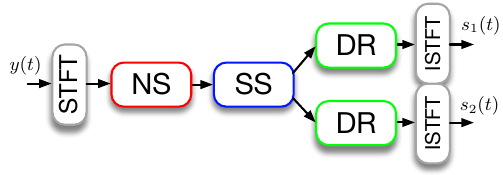}
    \caption{Cascaded model architecture. Modules for noise suppression (NS), reverberant speech separation (SS), and de-reverberation (DR) each contain the architecture shown in \cref{fig:cruse:2out}.}
    \label{fig:modular:architecture}
\end{figure}

The proposed cascaded model is represented in \cref{fig:modular:architecture} and involves three modules: 1) \acf{NS}, 2) \acf{SS}, and 3) \acf{DR}.
Splitting a system into a chain of task-specific modules was shown to be promising in \cite{Maciejewski:2020:Whamr:dataset:reverb}.

\begin{figure}[t!]
\centering
    \subfloat[Multi-decoder.]{
    \includegraphics[scale=.6]{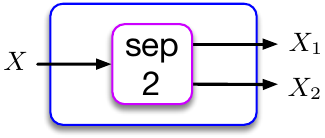}
  } \qquad
    \subfloat[\ac{SUB}.]{
    \includegraphics[scale=.6]{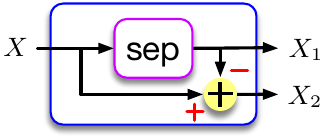}
  }
    \caption{Versions of separation modules (\ac{SS}) that separate a reverberant mixture $X$ into two reverberant sources $X_1$ and $X_2$.}
    \label{fig:sep:bar}
\end{figure}

We consider two configurations of the SS module, represented in \cref{fig:sep:bar}.
Configuration (a) has $D$ decoders, requiring prior information on the number of sources and using $D=R$.
It therefore jointly infers the separate speech signals from the input mixture.
Alternatively, we propose configuration (b) that has a single decoder, which outputs one speech estimate at a time.
This estimate is subtracted from the module's input to give a second output for the second speech estimate, and therefore termed \acf{SUB}. While the \ac{SUB} approach can be continued in a recursive fashion to generalize to an arbitrary number of sources, we restrict ourselves to $R=2$ in this work. 
The second configuration is more efficient in terms of memory and complexity as it only involves one decoder.
 

\subsection{End-to-end single stage model}

As a baseline and to evaluate the effectiveness of the cascaded model, we design a single end-to-end (E2E) system, which directly estimates the separated, de-reverberated, and noise suppressed speech signals. The E2E model therefore has to do the three tasks, NR, SS and DR, in a single step.
The E2E system is designed to have comparable complexity to the modularized version in Fig.~\ref{fig:sep:bar}, but consists only of one single module of the type shown in Fig.~\ref{fig:cruse:2out}.
Using the general architecture described in \cref{ssec:general:model}, we increase the capacity for the E2E by replacing the single GRU bottleneck by a three-layer LSTM. 
Moreover, we increase the temporal receptive field of the encoding and decoding convolutional networks with $(3,3)$ kernel sizes, still maintaining temporal causality.

\section{Loss functions}

We use utterance-level permutation invariant training (uPIT) to solve the source separation permutation problem \cite{2017:Kolbaek:uPIT}. In general, uPIT can be used with any distance metric, which is discussed in \ref{ssec:losses}. In \ref{ssec:softthresh}, an optional soft thresholding of the distance metric is discussed.


\subsection{Distance metric}
\label{ssec:losses}

As a baseline, we use the negative scale-invariant signal-to-distortion ratio (SI-SDR) \cite{Roux:2019:Si:SDR}, which is the \ac{SOTA} distance metric in DNN speech separation as used by ConvTasNet and SepFormer.

The complex compressed mean-squared error (CCMSE) \cite{Braun:2021:Loss:Function:Speech:Enhancement} has shown greater success in speech enhancement tasks targeting perceptual quality, and has also been shown in \cite{Aroudi:2021:Dbnet:Separation:Beamforming} to work well for speech separation. 
The CCMSE between target source $r$ and estimate $d$ is given as
\begin{multline}
\label{eqn:ccmse}
\mathcal{L}(\vect{S}_r, \widehat{\vect{S}}_d) = \frac{1}{KN} \sum_{k=1}^K \sum_{n=1}^N (1-\lambda)\Big \lvert \lvert S_r(k,n) \rvert^c - \lvert \widehat{S}_d(k,n) \rvert^c \Big \rvert ^ 2  \\
+ \lambda \Big \lvert  \lvert S_r(k,n) \rvert ^ c e^{\iu \Phi_{S_r(k,n)}} - \lvert \widehat{S}_d(k,n) \rvert ^ c e^{\iu \Phi_{\widehat{S}_d(k,n)}} \Big \rvert ^ 2 \, ,
\end{multline}
where $0 < c \leq 1$ is a compression factor, $0 \leq \lambda \leq 1$ is a weighting of magnitude-based and complex-valued errors, and $\Phi_S$ is the phase of $S \in \complexes$.
Note that CCMSE is equivalent to MSE when $c \!=\! \lambda \!=\! 1$. To remove the signal level dependence from the distance metric, the signals $S$ and $\widehat{S}$ are normalized by the active speech level as in \cite{Braun:2021:Loss:Function:Speech:Enhancement}.

\subsection{Soft thresholding}
\label{ssec:softthresh}
Using either of the aforementioned metrics, details of a louder talker will be more important than a quiet talker. In \cite{Wisdom:2020:Mixture:Invariant:Unsupervised,Landwehr:2021:Reverberant:Separation:Sepformer}, a soft threshold is added to the distance metric which limits it after a certain amount of separation is reached. If the louder talker's separation metric surpasses this threshold, the optimization shifts focus to reduce the estimation error of the quieter talker.
We adopt this approach by adding a soft threshold term $\tau \in \reals_{\geq 0}$ to the CCMSE, 
\begin{align}
    \mathcal{L}_\tau (\vect{S}_r, \widehat{\vect{S}}_d) = 10 \log_{10} \left( \mathcal{L}(\vect{S}_r, \widehat{\vect{S}}_d) + \tau \right) \, .
\end{align}
A fixed, signal-independent soft threshold $\tau$ is appropriate because our metric $\mathcal{L}$ in \cref{eqn:ccmse} is level-independent.

\section{Experimental procedures}
\label{sec:experiment:procedure}

\subsection{Datasets}

Considering that the performance of deep neural networks depends on the data used to train, validate, and evaluate the models, we carefully build realistic, large data sets to ensure the model generalizes to unseen data and tests well on real speech separation scenarios.
Training, validation and test data are drawn from different corpora and augmented through a diverse set of processes.

Training data is created \say{on-the-fly} by generating synthetic mixture signals of 10~s duration, which gives access to the input signal, its clean version, and the speech source signals in their reverberant, early reflection, and anechoic versions, as desired for task-specific training of the proposed modules.
The speech signals are concatenated from speech recordings that are publicly available in the 2021 Deep Noise Suppression (DNS) challenge \cite{Reddy:2021:DNS:Challenge} and AVspeech \cite{Ephrat:2018:AV:Speech:Dataset} datasets, and the same as used in our prior work  \cite{Braun:2021:Real:Time:Noise:Suppression:CRUSE}.
Non-reverberant speech signals were augmented with room impulse responses (RIRs) simulated with the image-source method \cite{Allen:1979:Image:Method:Acoustics}.
In total, the original source data includes 246 h noise, about 700 h speech, and 128k RIRs simulated in 2000 rooms.
The room size and reverb time varied between direct path and hall reverb.
Noise was added to the reverberant speech mixtures.
These noises included synthetic ones like filtered white noise, and real noises recorded from rooms and public spaces.
Signals were augmented through a variety of transformations like pitch shifting, spectral modifications like filtering, random silences, and variable number of active speakers in the mixture. Parameters that control these transformations were randomly sampled. 
A validation set of 300 sequences was generated using speech, noise, and RIRs from corpora that was different from the training data.

While speech separators are typically designed to output anechoic speech, the inclusion of early reflections can help with intelligibility \cite{Bradley:2003:Early:Reflections:Speech}, and can relax the enhancement problem.
To investigate how separation performance and speech quality are affected, we use either the direct-path (anechoic) speech signals or speech signals that include early reflections as training targets. The early reflection training target signals are obtained by applying a window to the RIR, keeping the first 50~ms after the direct sound.

While synthetically mixed datasets are commonly used to test speech separators, we advocate the testing of separators on real single-channel recordings to reflect real-world conditions.
Specifically, we tested the trained models with the REAL-M \cite{Subakan:2022:Real:M:Dataset} 
dataset, which contains 1436 real-world single-microphone recordings of two speakers. REAL-M is a crowd-sourced speech-separation corpus of two talkers recorded in different acoustic environments from a variety of devices like laptops and smartphones microphones. 

\subsection{Evaluation metrics}

The algorithms are evaluated in terms of speech quality and amount of separation, which are typically contradicting terms: Higher separation introduces more speech distortion.
For fair comparison with batch processing baselines on rather short test examples, the online processing algorithms are initialized by a prior pass on the test sequence to initialize its buffers and states to a converged state.
Since we evaluate on real recordings of overlapping speech, without access to ground truth sources, we use blind estimators of the metrics.

Speech quality is measured with DNSMOS \cite{Reddy:2020:DNSMOS}, a non-intrusive estimator of mean opinion score (MOS) in the context of noise suppression, providing scores for signal quality (SIG), background noise (BAK) and overall (OVR).
To estimate the separation quality from real recordings, we used the non-intrusive SI-SDR estimator \cite{Subakan:2022:Real:M:Dataset}, which was specifically trained for the REAL-M dataset. 

Further, we propose a novel non-intrusive metric, the \ac{CSE} given by 
\begin{equation}
\label{eq:cse}
    \text{CSE} = - 20 \log_{10}  \frac{|\widehat{\vect{s}}_r \tpose \widehat{\vect{s}}_{r'}| }{ \Vert\widehat{\vect{s}}_r\Vert^2 + \Vert\widehat{\vect{s}}_{r'}\Vert^2 } \, .
\end{equation}
By assuming constant source suppression factors $\alpha_r$ for separation algorithms, a model output for estimating source $s_r(t)$ is given by $\widehat{s}_r(t) = s_r(t) + \alpha_r s_{r'}(t) + \beta_r \eta(t)$. Given there is no correlation between source signals and noise, i.e.\,$\ev{\vect{s}_r \tpose \vect{s}_{r'}} = 0$, $\forall r\neq r'$,  and $\ev{ \vect{s}_r \tpose \vect{\eta}} = 0$, $\forall r$, we can interpret \eqref{eq:cse} as the normalized cross correlation between source $r$ and $r'$ by 
\begin{equation}
\label{eq:cse_derviation}
    \frac{\widehat{\vect{s}}_r \tpose \widehat{\vect{s}}_{r'} }{ \Vert\widehat{\vect{s}}_r\Vert^2 + \Vert\widehat{\vect{s}}_{r'}\Vert^2 } 
    = \frac{\alpha_r \Vert\vect{s}_r\Vert^2 + \alpha_{r'} \Vert \vect{s}_{r'}\Vert^2 }{ (1+\alpha_{r'}^2) \Vert\vect{s}_r\Vert^2 +  (1+\alpha_{r}^2) \Vert\vect{s}_{r'}\Vert^2 } \, .
\end{equation}
For $\alpha_{r}^2 \ll 1$, the term \eqref{eq:cse_derviation} becomes an unbiased estimate of the signal-power weighted average of the cross-talk factors $\alpha_r,\alpha_{r'}$, and can be seen as an estimator of channel separation. 


Finally, each model's separation and audio quality is related to its computational complexity at inference time according to multiply-accumulate (MAC) operations per 10~ms of audio.

\subsection{Algorithm parameters, model configurations, and baselines}

Audio data has a sampling frequency of 16 kHz.
For feature extraction, we use an STFT with 20~ms square-root-Hann windows and 50\% overlap.
The FFT size is set to $N_{\text{FFT}} = 320$.
The network input are the power-compressed 161 complex frequency components as two channels (real and imaginary part). 
The number of channels in the convolutional layers of the CRUSE modules varies depending on the task. Channels for \ac{NS} are 32-64-64-64 as in \cite{Braun:2021:Real:Time:Noise:Suppression:CRUSE}, whereas for \ac{SS} and \ac{DR} we find that a slightly more powerful model with a configuration of 32-64-128-256 helps performance. We use a GRU RNN for \ac{NS} and \ac{DR}, and an LSTM for \ac{SS}, as we find that the capacity of the recurrent network is a central factor that affects how well the network can separate reverberant speech.

The models are trained with a batch size of 128 sequences of 10 second duration using the AdamW \cite{Loshchilov:2019:AdamW} optimizer, learning rate of $1 \times 10^{-4}$, and weight decay of $2 \times 10^{-5}$. 
One training epoch is defined as 5120 sequences, and training is completed after around 5000 epochs, giving a pseudo-unique training data duration of about 8 years due to random data augmentation.
Modules are trained sequentially in order of their position in \cref{fig:modular:architecture}. First, \ac{NS} is trained to clean the noisy reverberant speech mixtures. Second, SS is trained to output separated reverberant speech signals using the output of the trained, frozen NS. Finally, the DR module is trained to de-reverberate the outputs of the trained, frozen SS and NS in cascade.

As an external baseline, we use the \ac{SOTA} large offline speech separation model SepFormer \cite{Subakan:2022:Transformer:Speech:Separation} from SpeechBrain \cite{speechbrain} that uses a pair of transformer networks to integrate temporal information.

\section{Results and discussions}
\label{sec:results}

\begin{figure}[t!]
    \centering
    \includegraphics[width=\linewidth,clip,trim=0 14 0 0]{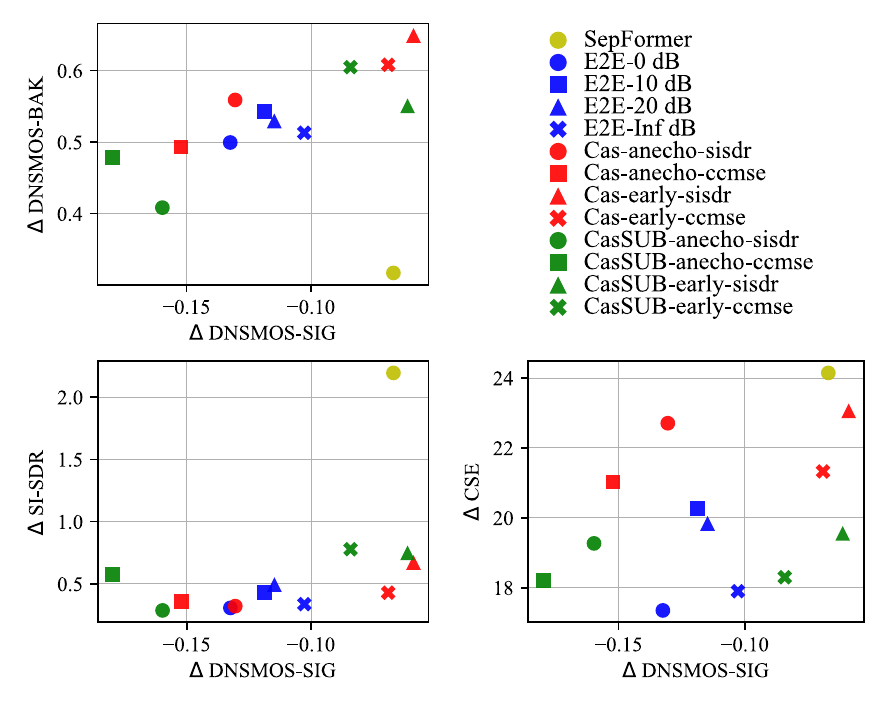}
    \vspace{-16pt}
    \caption{Overall results from REALM dataset evaluations.}
    \label{fig:results:realm}
\end{figure}

Overall results from testing the algorithms on the REAL-M dataset are shown in \cref{fig:results:realm} in terms of DNSMOS-SIG vs.\, DNSMOS-BAK, SI-SDR, and \ac{CSE}.
Qualitative results of an algorithm's output are expressed in terms of its improvement ($\Delta$) over the unprocessed input.
All of our proposed architectures are more efficient at inference time than the current \ac{SOTA} SepFormer architecture, with MACs per 10~ms of audio being an order of magnitude less: \textit{SepFormer} has 626 MACs, \textit{E2E} has 60 MACs, Cascade (\textit{Cas}) has 54 MACs, and Cascade Subtractive (\textit{CasSUB}) has 46 MACs.

Among the proposed models, the cascaded versions have the best efficiency (lowest cost) and performance in terms of all metrics. 
The cascaded versions with \ac{SUB} separation are the most efficient in terms of MACs because their separation module requires only one decoder.
While \ac{SUB} has a lower \ac{CSE} than the multi-decoder version, it has a similar DNSMOS-SIG and DNSMOS-BAK and higher SI-SDR.
This may be because the \ac{SUB}'s encoder and recurrent networks only need to represent the information of one source rather than two, so it may have more representation capacity than the multi-decoder version given the same latent dimensionality.
Training \ac{SUB} using a CCMSE loss with $\lambda=1$ and $c=.5$ gives higher SI-SDR and DNSMOS-BAK and lower DNSMOS-SIG and CSE than using an SI-SDR loss, coinciding with experiments in \cite{Aroudi:2021:Dbnet:Separation:Beamforming}.
\emph{Cas} has higher DNSMOS-BAK and SIG than \emph{SepFormer}.
Further, training on targets including early reflections give slightly better speech quality and separation quality scores than on anechoic targets.

\cref{table:stages:realm:ccmse} shows quantitative results from different stages of the cascaded system using a CCMSE loss. The \ac{NS} stage improves all DNSMOS metrics. The \ac{SS} stage introduces some artifacts compared to the original signal which lowers the DNSMOS-SIG. In terms of DNSMOS-BAK, each stage contributes some suppression of noise. In contrast to the anechoic DR module, the \ac{DR} module using early reflection training targets has better signal quality and channel separation. This suggests that the latter training target relax the problem and therefore lead to improved quality.

\begin{table}[t!]
    \centering
    \footnotesize
    \begin{tabular}{l|c|c|c|c}
    \toprule
    \multirow[c]{2}{*}{Stage} &   \multirow[c]{2}{*}{NS} &  \multirow[c]{2}{*}{NS + SS}  & \multicolumn{2}{c}{NS + SS + DR}  \\ 
    \cline{4-5}
    & & &  early &  anechoic \\
    \midrule
    $\Delta$ DNSMOS-SIG & 0.17 &  -0.08 &           -0.07 &              -0.15 \\
    $\Delta$ DNSMOS-BAK & 0.50 &   0.52 &            0.61 &               0.49 \\
    $\Delta$ DNSMOS-OVR & 0.32 &   0.11 &            0.17 &              -0.04 \\
    $\Delta$ CSE & 0.00 &  20.31 &           21.33 &              21.02 \\
    \bottomrule
\end{tabular}

    \caption{DNSMOS and channel separation improvement of output from each stage of a cascade system trained with CCMSE loss.}
    \label{table:stages:realm:ccmse}
\end{table}

\begin{figure}[t!]
    \centering
    \includegraphics[width=\linewidth,trim=0 10 0 10]{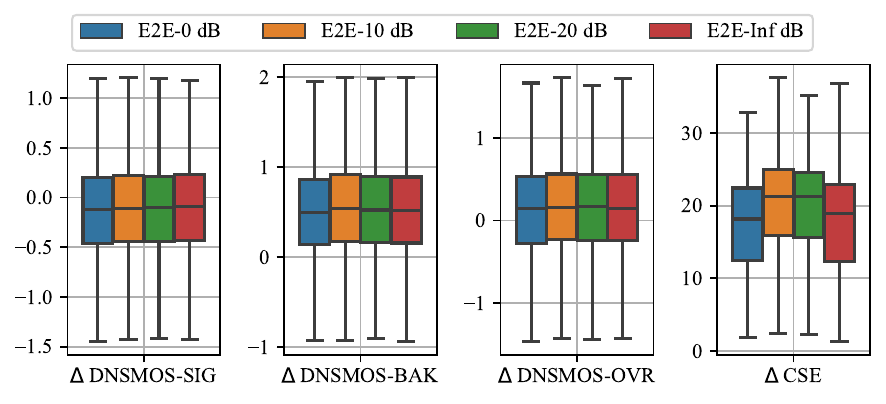}
    \caption{Result of using different soft thresholds $\tau$ in dB when training the E2E model using CCMSE loss on early reflection targets.}
    \label{fig:results:realm:experiments:thresh}
\end{figure}

Comparing the influence of soft threshold training on E2E models in \cref{fig:results:realm:experiments:thresh}, we observe that the highest DNSMOS-BAK, DNSMOS-OVR and CSE is obtained with a threshold of $-10$~dB, whereas the highest DNSMOS-SIG is obtained without a soft threshold (-Inf~dB).
Loss soft thresholding thus enables better separation in terms of correlation between sources at the cost of slightly lower averaged DNSMOS-SIG.


\section{Conclusions}
\label{sec:conclusions}

Encapsulating the tasks of \ac{NS}, \ac{SS}, and \ac{DR} into individual DNN
modules can achieve more efficient systems at better separation and speech quality than using a single end-to-end DNN model at comparable capacity for all three tasks. Our resource-efficient, causal, and compact architectures enable real-time separation of noisy and reverberant speech mixtures, which is shown on real-world single-channel recordings through the blind estimation of MOS and our proposed channel separation measure. Compared with large non-real-time \ac{SOTA} DNN models operating on raw waveforms, our real-time models separate time-frequency domain representations and use deep filters to give high quality separated signals. 
Training compact causal models on early reflection targets allows the system to output higher signal quality estimates as opposed to using anechoic targets. Resource efficiency is maximized when separation is actualized by our proposed \ac{SUB} separation module that uses only one decoder. Soft thresholding our CCMSE loss at -10 dB minimum gives better separation between sources, as it focuses on secondary talkers. Future work is required to improve the \ac{SUB} separation module to recursively disentangle an arbitrary number of talkers.

\vfill\pagebreak
\balance

\bibliographystyle{IEEEbib}
\bibliography{bibliography}

\begin{thebibliography}{10}

\bibitem{Darwin:2007:Speech:Segregation:Cocktail}
C.~J. Darwin,
\newblock ``Listening to speech in the presence of other sounds,''
\newblock {\em Philosophical Transactions of the Royal Society}, vol. 363, pp.
  1011--1021,  2007.

\bibitem{Wang:2018:Speech:Separation:DNN:Overview}
D. Wang and J. Chen,
\newblock ``Supervised speech separation based on deep learning: An overview,''
\newblock {\em IEEE/ACM Transactions on Audio, Speech, and Language
  Processing}, vol. 26, no. 10, pp. 1702–1726,  2018.

\bibitem{Luo:2019:ConvTasNet}
Y. Luo and N. Mesgarani,
\newblock ``Conv-{T}as{N}et: Surpassing ideal time-frequency magnitude masking
  for speech separation,''
\newblock {\em IEEE Trans. Audio, Speech, Lang. Process.}, vol. 27, no. 8, pp.
  1256--1266,  2019.

\bibitem{Luo:2020:Dual:Path:Rnn:Speech:Separation}
Y. Luo, Z. Chen, and T. Yoshioka,
\newblock ``Dual-path {RNN}: efficient long sequence modeling for time-domain
  single-channel speech separation,''
\newblock in {\em Proceedings of the IEEE International Conference on
  Acoustics, Speech and Signal Processing (ICASSP)}. IEEE,  2020, pp. 46--50.

\bibitem{Subakan:2022:Transformer:Speech:Separation}
C. Subakan, M. Ravanelli, S. Cornell, M. Bronzi, and J. Zhong,
\newblock ``Attention is all you need in speech separation,''
\newblock in {\em Proceedings of the IEEE International Conference on
  Acoustics, Speech and Signal Processing (ICASSP)},  2021, pp. 21--25.

\bibitem{Vincent:2018:Source:Separation}
E. Vincent, T. Virtanen, and S. Gannot, Eds.,
\newblock {\em Audio Source Separation and Speech Enhancement},
\newblock John Wiley and Sons,  2018.

\bibitem{Maciejewski:2020:Whamr:dataset:reverb}
M. Maciejewski, G. Wichern, E. McQuinn, and J. Le~Roux,
\newblock ``{WHAMR}!: Noisy and reverberant single-channel speech separation,''
\newblock in {\em Proceedings of the IEEE International Conference on
  Acoustics, Speech and Signal Processing (ICASSP)}. IEEE,  2020, pp. 696--700.

\bibitem{Landwehr:2021:Reverberant:Separation:Sepformer}
T. Cord-Landwehr, C. Boeddeker, T. Von~Neumann, C. Zorilă, R. Doddipatla, and
  R. Haeb-Umbach,
\newblock ``Monaural source separation: From anechoic to reverberant
  environments,''
\newblock in {\em 2022 International Workshop on Acoustic Signal Enhancement
  (IWAENC)},  2022, pp. 1--5.

\bibitem{2018:Kinoshita:Subtractive:Recurrent:Hearing:Nets}
K. Kinoshita, L. Drude, M. Delcroix, and T. Nakatani,
\newblock ``Listening to each speaker one by one with recurrent selective
  hearing networks,''
\newblock in {\em IEEE International Conference on Acoustics, Speech and Signal
  Processing (ICASSP)},  2018, pp. 5064--5068.

\bibitem{Takahashi:2019:Recursive:Speech:Separation:unknown}
N. Takahashi, S. Parthasaarathy, N. Goswami, and Y. Mitsufuji,
\newblock ``Recursive speech separation for unknown number of speakers,''
\newblock in {\em Interspeech},  2019.

\bibitem{Wisdom:2020:Mixture:Invariant:Unsupervised}
S. Wisdom, E. Tzinis, H. Erdogan, J. Weiss, K. Wilson, and J. Hershey,
\newblock ``Unsupervised sound separation using mixture invariant training,''
\newblock {\em Advances in Neural Info. Process. Systems},  2020.

\bibitem{Roux:2019:Si:SDR}
J.~L. Roux, S. Wisdom, H. Erdogan, and J.~R. Hershey,
\newblock ``{SDR} – half-baked or well done?,''
\newblock in {\em Proceedings of the IEEE International Conference on
  Acoustics, Speech and Signal Processing (ICASSP)},  2019, pp. 626--630.

\bibitem{Braun:2021:Real:Time:Noise:Suppression:CRUSE}
S. Braun, H. Gamper, C.~K. Reddy, and I. Tashev,
\newblock ``Towards efficient models for real-time deep noise suppression,''
\newblock in {\em Proceedings of the IEEE International Conference on
  Acoustics, Speech and Signal Processing (ICASSP)},  2021, pp. 656--660.

\bibitem{Talmon:2009:CTF:Transfer:Function}
R. Talmon, I. Cohen, and S. Gannot,
\newblock ``Multichannel speech enhancement using convolutive transfer function
  approximation in reverberant environments,''
\newblock in {\em 2009 IEEE International Conference on Acoustics, Speech and
  Signal Processing},  2009, pp. 3885--3888.

\bibitem{Mack:2020:Deep:Filtering}
W. Mack and E.~A.~P. Habets,
\newblock ``Deep filtering: Signal extraction and reconstruction using complex
  time-frequency filters,''
\newblock {\em IEEE Signal Processing Letters}, vol. 27, pp. 61--65,  2020.

\bibitem{2017:Kolbaek:uPIT}
M. Kolbæk, D. Yu, Z.-H. Tan, and J. Jensen,
\newblock ``Multitalker speech separation with utterance-level permutation
  invariant training of deep recurrent neural networks,''
\newblock {\em IEEE/ACM Transactions on Audio, Speech, and Language
  Processing}, vol. 25, no. 10, pp. 1901--1913,  2017.

\bibitem{Braun:2021:Loss:Function:Speech:Enhancement}
S. Braun and I. Tashev,
\newblock ``A consolidated view of loss functions for supervised deep
  learning-based speech enhancement,''
\newblock in {\em 2021 44th International Conference on Telecommunications and
  Signal Processing (TSP)},  2021, pp. 72--76.

\bibitem{Aroudi:2021:Dbnet:Separation:Beamforming}
A. Aroudi and S. Braun,
\newblock ``{DB}net: {DOA}-driven beamforming network for end-to-end
  reverberant sound source separation,''
\newblock in {\em Proceedings of the IEEE International Conference on
  Acoustics, Speech and Signal Processing (ICASSP)},  2021, pp. 211--215.

\bibitem{Reddy:2021:DNS:Challenge}
C.~K.~A. Reddy, H. Dubey, V. Gopal, R. Cutler, S. Braun, H. Gamper, R. Aichner,
  and S. Srinivasan,
\newblock ``{ICASSP} 2021 deep noise suppression challenge,''
\newblock in {\em ICASSP 2021 - 2021 IEEE International Conference on
  Acoustics, Speech and Signal Processing (ICASSP)},  2021, pp. 6623--6627.

\bibitem{Ephrat:2018:AV:Speech:Dataset}
A. Ephrat, I. Mosseri, O. Lang, T. Dekel, K. Wilson, A. Hassidim, W.~T.
  Freeman, and M. Rubinstein,
\newblock ``Looking to listen at the cocktail party: A speaker-independent
  audio-visual model for speech separation,''
\newblock {\em ACM Trans. Graph.}, vol. 37, no. 4,  2018.

\bibitem{Allen:1979:Image:Method:Acoustics}
J.~B. Allen and D.~A. Berkely,
\newblock ``Image method for efficiently simulating small-room acoustics,''
\newblock {\em J. Acoust. Soc. Am.}, vol. 64, no. 4, pp. 943--950,  1979.

\bibitem{Bradley:2003:Early:Reflections:Speech}
J.~S. Bradley, H. Sato, and M. Picard,
\newblock ``On the importance of early reflections for speech in rooms,''
\newblock {\em The Journal of the Acoustical Society of America}, vol. 113, no.
  6, pp. 3233--3244,  2003.

\bibitem{Subakan:2022:Real:M:Dataset}
C. Subakan, M. Ravanelli, S. Cornell, and F. Grondin,
\newblock ``{REAL-M}: Towards speech separation on real mixtures,''
\newblock in {\em Proceedings of the IEEE International Conference on
  Acoustics, Speech and Signal Processing (ICASSP)}. IEEE,  2022, pp.
  6862--6866.

\bibitem{Reddy:2020:DNSMOS}
C. K.~A.~Reddy, V. Gopal, and R. Cutler,
\newblock ``{DNSMOS}: A non-intrusive perceptual objective speech quality
  metric to evaluate noise suppressors,''
\newblock in {\em Proceedings of the IEEE International Conference on
  Acoustics, Speech and Signal Processing (ICASSP)}.  2020, pp. 6493--6497,
  IEEE.

\bibitem{Loshchilov:2019:AdamW}
I. Loshchilov and F. Hutter,
\newblock ``Decoupled weight decay regularization,''
\newblock in {\em International Conference on Learning Representations (ICLR)},
   2019.

\bibitem{speechbrain}
M. Ravanelli, T. Parcollet, P. Plantinga, A. Rouhe, S. Cornell, L. Lugosch, C.
  Subakan, N. Dawalatabad, A. Heba, J. Zhong, J.-C. Chou, S.-L. Yeh, S.-W. Fu,
  C.-F. Liao, E. Rastorgueva, F. Grondin, W. Aris, H. Na, Y. Gao, R.~D. Mori,
  and Y. Bengio,
\newblock ``{SpeechBrain}: A general-purpose speech toolkit,''  2021,
\newblock arXiv:2106.04624.

\end{thebibliography}

\end{document}